# Multi-phonon Raman scattering in GaN nanowires

S. Dhara,[a),d] Sharat Chandra,[a] G. Mangamma, S. Kalavathi, P. Shankar, K. G. M. Nair,

and A. K. Tyagi

Materials Science Division, Indira Gandhi Center for Atomic Research, Kalpakkam-603102

C. W. Hsu,[a),b] C. C. Kuo, L. C. Chen and K. H. Chen [b]

Centre for Condensed Matter Sciences, National Taiwan University, Taipei-106, Taiwan

K. K. Sriram[c]

Department of Physics, PSG College of Technology, Anna University, Coimbatore -641004

*Abstract*

UV Raman scattering studies show longitudinal optical (LO) mode up to 4$^{th}$ order in wurtzite GaN nanowire system. Fröhlich interaction of electron with the long range electrostatic field of ionic bonded GaN gives rise to enhancement in LO phonon modes. Good crystalline quality, as indicated by the crystallographic as well as luminescence studies, is thought to be responsible for this significant observation. Calculated size dependence, incorporating size corrected dielectric constants, of electron-phonon interaction energy agrees well with measured values and also predict stronger interaction energy than that of the bulk for diameter below ~3 nm.

[a] These three authors contributed equally

[b] Also affiliated with Institute of Atomic and Molecular Sciences, Academia Sinica, Taipei 106, Taiwan

[c] Now at Institute for Atomic and Molecular Sciences, Academia Sinica, Taipei-106, Taiwan

[d] Author to whom all correspondence should be addressed; electronic mail: dhara@igcar.gov.in

GaN continues to be an important material due to its potential applications as white light source and UV diode. Wurtzite GaN is having a band gap ~3.47 eV in the ultraviolet (UV) region and is stable at high temperature and applicable in high power electronics.[1] Among several techniques, formation of one-dimensional (1-D) GaN nanowire (NW) is mostly reported in the vapor-liquid-solid (VLS) process where either of noble- or transition-metals,[2-4] and Ga based catalysts[5] are used to nucleate 1-D structure with the fluctuation of metastable liquid alloy phases formed with Ga vapor and finally solidifying at the alloy interface after reacting with $NH_3$. Optical properties are studied for these GaN NWs, extensively, starting from reflectance and transmission studies in the UV-visible-infrared region,[6,7] photoluminescence (PL),[2,8-10] and Raman scattering studies.[2,11-13] Plasma frequency ~1100 (120) $cm^{-1}$ for NW is reported from reflectance minima.[6] In PL studies, quantum confinement of encapsulated 1-D NW is observed for direct band-to-band transition peak.[8] Lasing is also realized in the single GaN NW.[9] On the other hand, yellow luminescence (YL) band ~ 2.2 eV is one of the most well discussed defect bands present in GaN film with either native point-defects[14] or point-defects nucleating at extended defects like dislocations as the origin.[15] Lately, there are reports of PL spectra without any YL band in 1-D NW system.[9,12] This is due to the fact that defect formation in nanocrystalline system is relatively difficult than that for buk.[16]

Extensive reports exist on Raman studies in NW system with sub-bandgap excitations.[2,11] Raman studies with UV excitation are also reported[12] with multi-phonon spectra restricted to $2^{nd}$ order longitudinal optical $A_1(LO)$ mode reported only for nanocrystalline GaN.[13] Fröhlich interaction arises due to coupling between LO phonon mode with available electron carriers in the excitation above the bandgap value. Eventually, it is responsible for a strong rise in first order LO mode, as well as, higher order modes corresponding to the primary one.[16] Detailed multi-phonon properties are reported for 2-D system of GaN film in resonant condition,[18] while there is hardly any report for 1-D GaN NWs.

We report here multi-phonon spectra for the GaN NWs. Morphological, crystallographic structural and optical properties with UV excitation are studied to characterize the grown nanostructures. Calculation on size dependence of electron-phonon interaction energy is validated successfully with experimental values for a large size range. Interaction energy is calculated in details with the incorporation of size corrected dielectric constants.

GaN NWs are grown on Au-coated crystalline (*c*-) Si(100) and amorphous (*a*-) SiO$_2$ substrates using Ga as source material and 10 sccm of NH$_3$ flow as reactant gas in a tubular furnace at atmospheric pressure by CVD technique. Substrate temperature is kept ~ 1173 K. Substrates are pre-coated with Au catalyst deposited by DC sputtering technique. Morphological and structural studies are performed using high resolution scanning electron microscope (HRSEM) and glancing incidence x-ray diffraction (GIXRD) system, respectively. Studies related to optical properties are performed using excitation of 325 nm line of He-Cd continuous wave laser and dispersion with three set of holographic blazed 1800 gr/mm gratings in a double subtractive triple monochromator. The experiments are performed in the back scattering configuration using a liquid N$_2$ cooled 'back-thinned' CCD detector for the detection of scattered intensity.

Straight NWs of micron meter in length are observed typically in HRSEM studies (Fig. 1) with average diameters, $D$ ~ 50 nm and 75 nm for NWs grown on *a*-SiO$_2$ and *c*-Si substrates, respectively. Crystallographic structural studies using GIXRD show formation of GaN in hexagonal (*h*-) wurtzite phase (Fig. 1). Peaks at 2θ values of 32.42, 34.61, 36.7, 48.3,

57.97, 63.7 and 69.25 correspond to crystalline planes (100), (002), (101), (102), (110), (103), and (112) of $h$-GaN (JCPDS : 02-1078). All major peaks reported in the JCPDS data for $h$-GaN are observed in our sample showing excellent crystalline quality of the grown NW. A rise in the background at lower angles, observed for the samples grown on $a$-SiO$_2$, is due to the amorphous nature of the substrate.

UV Raman study with excitation wavelength of 325 nm shows strong first order A$_1$(LO) at 727 cm$^{-1}$ (Fig. 2) along with progressively diminishing higher order modes 2A$_1$(LO) at ~1468 cm$^{-1}$, 3A$_1$(LO) at ~2197 cm$^{-1}$ and 4A$_1$(LO) at ~2930 cm$^{-1}$. In multi-phonon processes the phonons are no longer restricted to the center of the Brillouin zone and, therefore, the frequency of the higher order LO modes are slightly lower than the first order mode frequency multiplied by order.[19] The coupling with LO mode includes contributions from both the Fröhlich interaction that involves the long-range interaction generated by the macroscopic electric field associated with the LO phonons and the deformation potential. On the other hand, the transverse optic (TO) mode is mainly dominated by the deformation potential that involves the short-range interaction between the lattice displacement and the electrons. Absence of A$_1$(TO) mode at ~537 cm$^{-1}$ and it's higher order, corresponding to $h$-GaN (Fig. 2), rules out the role of deformation potential in the sample. Thus observation of the 4$^{th}$ order A$_1$(LO) mode ~2930 cm$^{-1}$ for NWs grown in $c$-Si substrate (inset in Fig. 3) indicates excellent crystalline quality in the grown sample.[18] Hardly any prominent 4$^{th}$ order peak is observed for NWs grown on $a$-SiO$_2$ substrate as it may be overshadowed by larger luminescence from the substrate around that region of the spectra (please see supplementary information Fig. S1). Inset in figure 2 show the fitted PL spectra (devoid of large contributions from $a$-SiO$_2$ substrate and analogous contribution from possible the SiO$_x$ layer in case of Si substrate) for NWs grown on $c$-Si and $a$-SiO$_2$ substrates with stronger signal in case of later. As a matter of fact, observation of higher order Raman modes become difficult in the riding PL signal of large intensity as discussed below.

Typical PL spectra with peaks corresponding to direct band gap ~3.46 eV (Fig. 3) are shown for the GaN NWs with $D \sim$ 20 nm to 40 nm grown with different conditions and detailed elsewhere.[2,3] using Au and Ni catalysts (3-4 nm deposited by electron beam evaporation technique), respectively. The spectra for the NWs are broadened with respect to that recorded for epitaxial (epi-) GaN film (Fig. 3). Among other reasons, 1-D nature in NWs may be responsible for it. Minor presence of peak corresponding YL band (~2.2 eV) in the epi-GaN may be due to the presence of dislocations (density <$10^8$ cm$^{-2}$) in the interface layer with (0001) Al$_2$O$_3$ substrate.[15] YL band is completely absent in case of the GaN NWs studied in this report, as similar observation is also made for NWs with bigger $D$ values (fitted PL data inset Fig. 2). Multi-phonon properties of these samples are also studied (inset Fig. 3). The good crystalline quality of grown NWs, as revealed by the absence of YL band in the PL spectra (inset Fig. 2 and Fig. 3), ensures a predominant Fröhlich interactions with higher order Raman modes being recorded in our samples.

We have calculated the Frank-Condon interaction energy ($E_{FC}$),[20] as a measure of the strength of the electron-phonon coupling for nanostructures of different $D$ values and validate with the experimental data. For the detailed calculation, we have also incorporated the variation in the static dielectric constant, $\varepsilon_0$, and the high frequency dielectric constant, $\varepsilon_\infty$, of the nanostructure as a function of the $D$. These have been calculated using the modified Penn model approach[21-24] and the Lyddane-Sachs-Teller (LST) relation[20] between the observed LO and TO mode frequencies in GaN, $\omega_{TO}$ and $\omega_{LO}$, and $\varepsilon_0$ and $\varepsilon_\infty$. Using the modified Penn model we can write[21,22]

$$\varepsilon_\infty^{NS} = 1 + \frac{\varepsilon_\infty^{bulk} - 1}{1 + (2\xi/D)^\gamma} \qquad (1)$$

Here, $\xi$ is a material dependent constant and $\varepsilon_\infty^{bulk}$ is the bulk high frequency dielectric constant. The exponent $\gamma$ is found to be material dependent in several approaches,[23,24] and varies between 1.2 and 2, but it can be simply taken to be 2 (Ref. 24). The $\xi$ parameter is calculated using the expression, $\xi = (\pi\hbar/E_g)\sqrt{E_2/2m}$, where, $E_2$ is the high energy interband transition in the bulk, $E_g$ is the fundamental band gap, $m$ is the rest mass of the electron and $\hbar$ the reduced Plank constant.[22] Once $\varepsilon_\infty^{NS}$ is known, we can calculate the $\varepsilon_0^{NS}$ from the LST relation as[20]

$$\frac{\omega_{LO}^2}{\omega_{TO}^2} = \frac{\varepsilon_0^{NS}}{\varepsilon_\infty^{NS}} \qquad (2)$$

As no TO mode is observed in our case, we take the LO and TO mode frequencies observed in the bulk GaN. This is justified given the large size of the nanowires studied and also the fact that the structure of the nanowires corresponds to that of the bulk GaN. Moreover, the reported values of $\omega_{TO}$ and $\omega_{LO}$ are almost equal to that observed in the bulk for NWs with large $D$ values.[2] For bulk GaN the various parameters are $E_2$ = 7.65 eV, $E_g$ = 3.47 eV, $\varepsilon_0^{bulk} = 10.4$ and $\varepsilon_\infty^{bulk} = 5.8$ (Ref. 25), where the dielectric constants are in the crystallographic $c$-direction. Using these parameter values we can calculate $\xi$ = 0.493 nm. Knowing the static and high frequency dielectric constants for the GaN nanostructure, we can calculate the Frank-Condon interaction energy for the nanostructures, $E_{FC}^{NS}$ as[20]

$$E_{FC}^{NS} = \frac{e^2}{2\pi D \epsilon_0} \left( \frac{1}{\varepsilon_\infty^{NS}} - \frac{1}{\varepsilon_0^{NS}} \right) \int_0^\pi \frac{(2u - \sin(2u))^2}{8\pi u^2} du \qquad (3)$$

Here, $\epsilon_0$ is the permittivity of the vacuum and $u$ is the dummy parameter for integration. These quantities can be easily calculated taking all the parameters in the SI units. The $E_{FC}$ for the bulk can be calculated from the relation[20]

$$E_{FC}^{bulk} = \frac{e^2}{8\pi \epsilon_0} \left( \frac{2m^* \omega_{LO}}{\hbar} \right)^{1/2} \left( \frac{1}{\varepsilon_\infty^{bulk}} - \frac{1}{\varepsilon_0^{bulk}} \right) \qquad (4)$$

Considering the effective mass $m^* = 0.22m$ for bulk GaN,[25] the bulk $E_{FC}$ value is calculated to be 37.975 meV (marked with arrow in Fig. 4). The variation of $E_{FC}^{NS}$ as a function of $D$ for nanowires is shown in Fig. 4. It is seen that the interaction energy decreases with the increasing $D$ values of the GaN NWs. An idea of the strength of interaction can be also be obtained from the experimental data of 2A$_1$(LO)/A$_1$(LO) intensity ratios. The measured values are also plotted as scattered points for various $D$ values of NWs in the range of ~20-75 nm (Fig. 4). Typical SEM images show (insets Fig. 4) NWs with $D \sim$ 30 nm and 40 nm. HRSEM images of NWs with bigger $D$ values are already shown in figure 1. The experimentally measured values show a similar trend as the calculated ones for a large size range. We must emphasize here that there are hardly any experimental evidence for the empirically calculated size dependence of $E_{FC}$ in nanostructures with a fairly large size range. As a matter of fact, a recent attempt with ZnO quantum dots is also futile.[26] Presence of impurity in the system, derived from carbonaceous (acetate) precursors may be the prime reason for not obtaining the desired result. Presence of impurity is prominent for the reported ZnO quantum dots with the presence of defect luminescence band and multi-phonon spectra being not so prominent.

The 2A$_1$(LO)/A$_1$(LO) intensity ratio corresponding to epi-GaN film is also shown as 'X' mark (Fig. 4). The electronic density of states in the epitaxial film and nanowires is different from that in the bulk thus resulting in a modification of the transition probability matrix elements in the nanostructures. This can explain the fact that even for the larger nanostructures we observe behavior different from that of the bulk.[27] However, our calculation indicates an interesting phenomenon that the electron-phonon interaction energy for nanostructures with $D$ value below ~3 nm can go stronger than that for the bulk value.

Efforts are being made to provide experimental evidence for GaN NWs with $D$ values close to it.

In conclusion, multi-phonon LO modes up to, as high as, $4^{th}$ order are recorded successfully for GaN nanowires with relatively low PL intensity. UV-Raman scattering in case of wide band gap semiconductors with ionic bonds, invoking long range Fröhlich interaction, is an excellent tool for identifying crystalline quality in the sample. Size dependent electron-phonon interaction energy is also calculated for nanostructured GaN system with necessary correction to dielectric constants in small dimensions. The interaction energy is found to increase with decreasing diameter of the nanowires and corroborates well with experimental values for a fairly large size range.

We thank D. Bhattacharjee of Department of Materials Science, Anna University, Guindy Campus, Chennai for her role as a project student. We also acknowledge J. C. George, V. Baskaran, B. K. Panigrahi, and S. Dash of MSD, IGCAR for extending their help in setting up the growth chamber. We thank V. S. Sastry of MSD, IGCAR, for his help in GIXRD measurements. We thank V. Subramanya Sarma of IIT Madras, Chennai for HRSEM studies. Finally, we acknowledge C. S. Sundar of MSD, IGCAR for his illuminating discussion during the preparation of the manuscript.

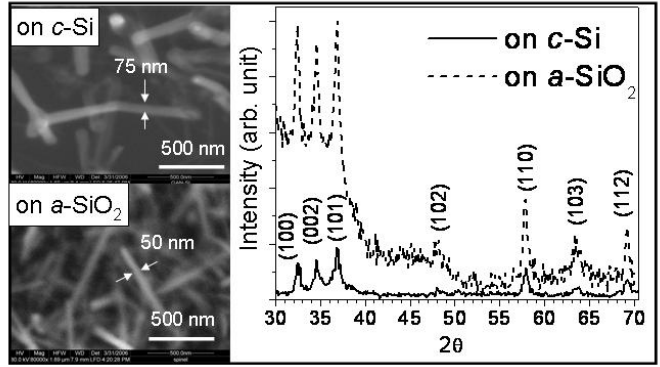

Fig.1. HRSEM images of nanowires grown on *c*-Si and *a*-SiO$_2$ substrates. Crystallographic structural studies using GIXRD show wurtzite GaN phase grown on both the substrates.

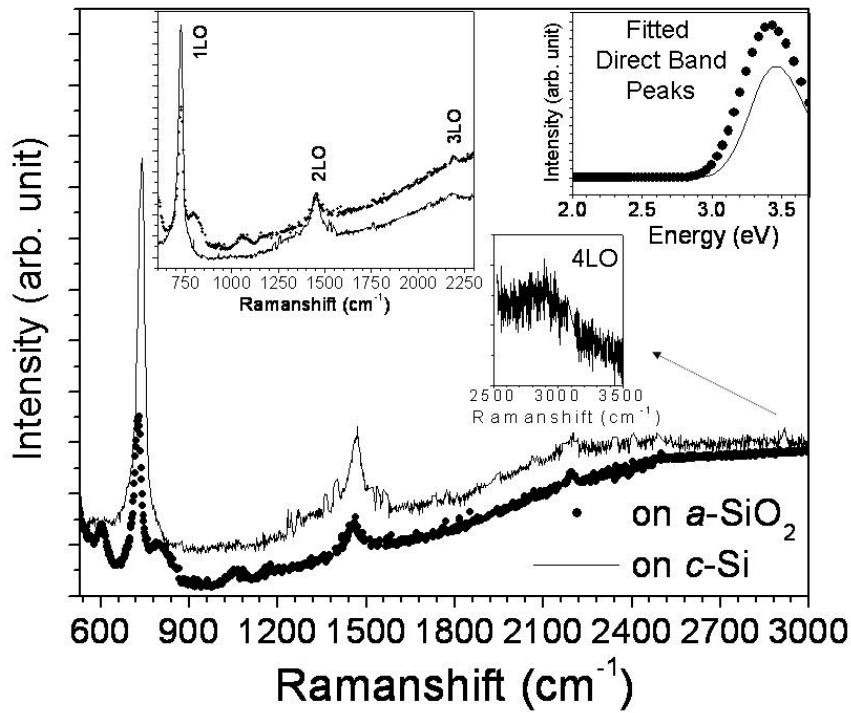

Fig. 2. UV Raman spectra showing higher order A$_1$(LO) modes for GaN nanowires grown on *c*-Si and *a*-SiO$_2$ substrates. Insets show slow scan upto 3$^{rd}$ order modes for samples grown on both the substrates. A clear 4$^{th}$ order mode is recorded for sample grown on *c*-Si. Fitted PL peak ~3.46 eV corresponding to direct band gap is also shown in the inset.

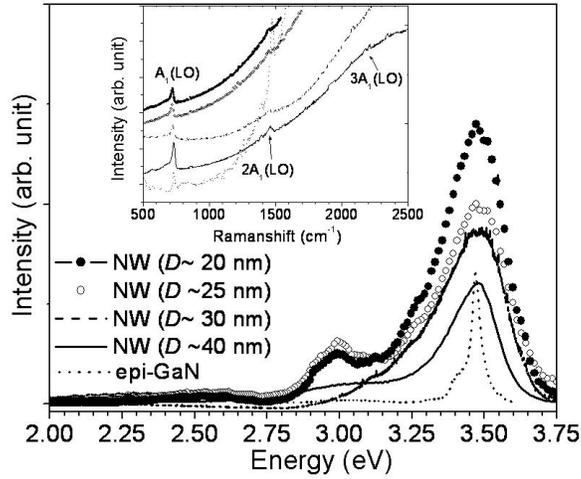

Fig.3. PL spectra of the epi-GaN and NWs with average diameter, $D$ ~20 nm to 40 nm. Epi-GaN shows a minor YL band along with a sharp direct band peak. A relatively broad direct band peak is shown for NWs with the absence of YL band. Inset shows multi-phonon properties of these samples. Large backgrounds indicate strong PL signal in the samples.

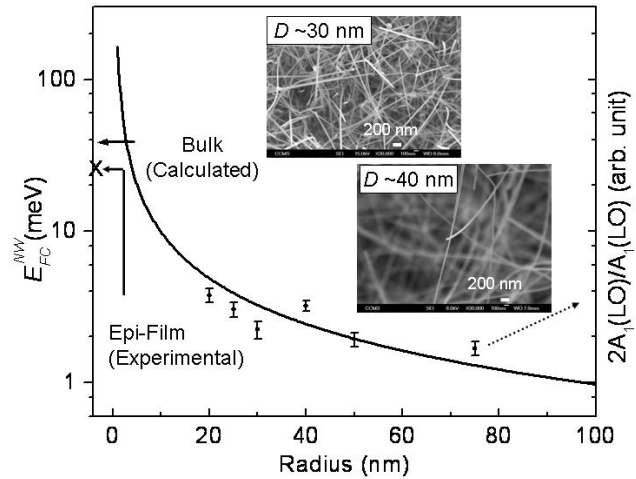

Fig. 4. Size dependent plot for calculated (continuous line) Frank-Condon interaction energy ($E_{FC}^{NS}$) as well as experimentally observed 2A$_1$(LO)/A$_1$(LO) values (scattered points with error bar) for GaN nanostructures. Calculated value for bulk GaN and experimental data for epi-GaN is also marked in the figure. Inset shows HRSEM images of NWs with average diameter, $D$ ~ 30 nm and 40 nm.

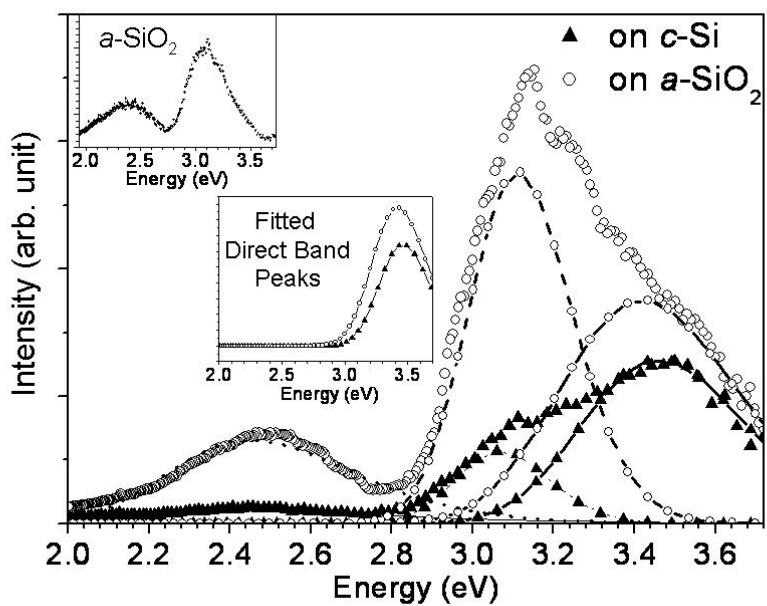

Fig. S1. Photoluminescence spectra for GaN nanowires grown on *c*-Si and *a*-SiO$_2$ substrates showing broad peak at high energy. A Gaussian peak fitting shows contribution from peaks ~ 3.46 eV, 3.06 eV and 2.48 eV. Inset shows the PL spectrum for *a*-SiO$_2$ substrate. Peaks at ~ 3.06 eV (dashed lines with respective symbols) and at ~2.48 eV (dotted lines) are contributions from the substrate. Fitted peaks ~3.46 eV (continuous lines with respective symbols; also shown as one of the insets in Fig 2) corresponding to direct band gap is also shown separately in the inset.